%% file: main.tex
\documentclass[reprint,aps,prl,twocolumn,floatfix,secnumarabic,hidelinks,
superscriptaddress]{revtex4-2}

\usepackage[final]{changes}
\setauthormarkuptext{}
\definechangesauthor[name={Referee 1}, color=red]{R1}
\definechangesauthor[name={Referee 2}, color=blue]{R2}
\definechangesauthor[name={Own Changes}, color=olive]{OC}

\input{style}

\input{commands}

\begin{document}
\title{\Title{}}

\input{title-config}
\begin{abstract}\noindent
Building fault-tolerant quantum computers with large numbers of logical qubits requires both scalable hardware architectures and error-correcting codes that make efficient use of physical qubits.
Photonic interconnects address both of these challenges by allowing the physical qubits to be distributed across multiple processors~\cite{grover1997quantum, cirac1999distributed}, while providing the non-local connectivity necessary to implement resource-efficient codes such as high-rate \ac{qLDPC} codes~\cite{breuckmann2021quantum, bhardwaj2026highrate}.
A key requirement for realising this architecture is the ability to perform stabiliser measurements between remote processors, which has not previously been demonstrated experimentally.
Here we report the first experimental demonstration of distributed quantum error detection and correction.
We generate entanglement between \network{} qubits in two separate trapped-ion processors and use it to perform remote syndrome measurements on \data{} qubits.
We first realise a distributed $[[2,1,1]]$ repetition code, detecting phase-flip errors on a logical qubit encoded across the two modules in real time and suppressing logical errors. 
We then combine these mid-circuit syndrome measurements with real-time feedforward to actively correct arbitrary single-qubit Pauli errors on a distributed Bell state. 
These results provide an experimental foundation for \ac{QEC} across modular quantum architectures.

\acresetall
\end{abstract}

\maketitle
\noindent

\section{Introduction}
Scaling quantum computers to the large numbers of physical qubits required for fault-tolerant computation remains a major hardware challenge.
Modular architectures offer a route beyond the capacity of individual processors, distributing the required physical qubits across multiple devices operating as a single computational machine~\cite{grover1997quantum,cirac1999distributed,monroe2014largescale}.
Even with such modular scaling, however, the size of this machine depends critically on the physical-qubit overhead of the error-correcting code.
High-rate \ac{qLDPC} codes offer substantially improved encoding efficiency compared to two-dimensional topological codes such as the surface code~\cite{breuckmann2021quantum,cain2026shors, bhardwaj2026highrate, webster2026pinnacle}, with recent proposals achieving encoding rates exceeding $1/2$ at practically relevant scales~\cite{kasai2026breaking, lu2026quantum, yang2026designer, zhao2026towards}.
This improved encoding efficiency comes at the cost of connectivity, as although each stabiliser acts on only a small number of qubits, the interactions required to measure them are generally not restricted to qubits that are nearby in the physical device.
\begin{figure*}[!t]
    \centering
    \includegraphics[width=178mm]{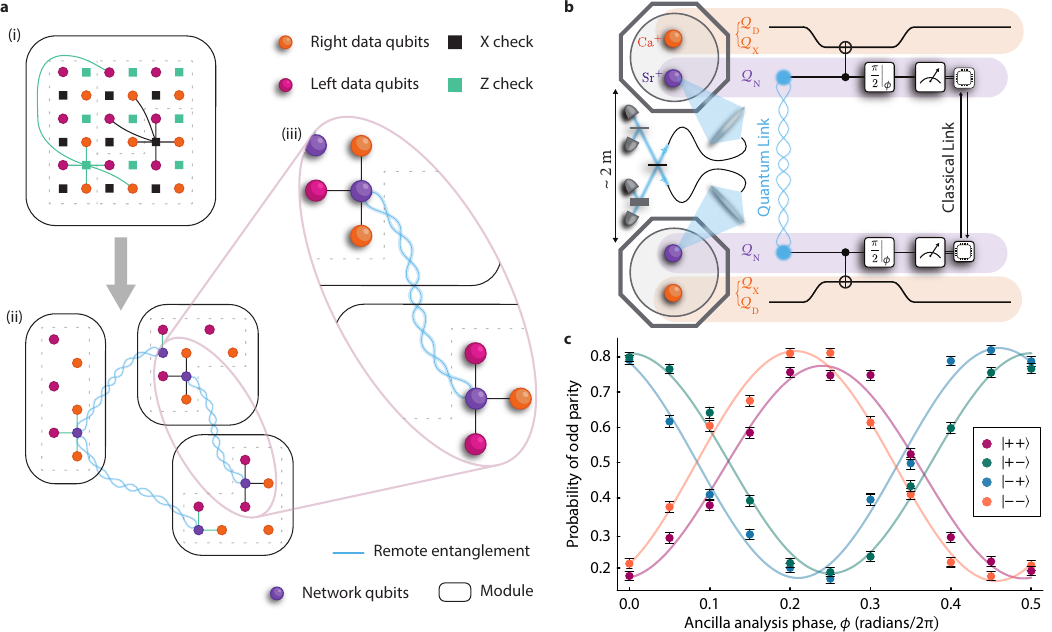}
    \caption{%
    \justifying
    \textbf{Remote stabiliser measurements.}
    \textbf{a}, Illustrative example of remote stabiliser measurements in a modular quantum architecture. 
    \textbf{i}, Tanner graph of the $[[18,4,4]]$ bivariate bicycle code~\cite{bravyi2024high, wang2026demonstration}, comprising 18 \data{} qubits divided into left and right sets, together with $X$- and $Z$-check nodes. Each stabiliser acts on six \data{} qubits, but the code requires interactions between qubits that cannot all be made local in a two-dimensional device layout.
    \textbf{ii}, The code can instead be distributed across three modules, each containing six \data{} qubits, reducing both the number of qubits and the required local connectivity within each processor.
    \textbf{iii}, Inter-module stabiliser measurements are mediated by photonic interconnects. Entanglement between \network{} qubits provides the shared resource required to perform these measurements across separate processors.
    \textbf{b}, Experimental implementation of a two-qubit remote stabiliser measurement. Two modules each hold a \Sr{88} ion providing a \network{} qubit, $\Qnet$, and a \Ca{43} ion providing a \data{} qubit, $\Qdat$, and auxiliary qubit, $\Qaux$. A Bell pair shared between the \network{} qubits provides a distributed ancilla for measuring remote \data{}-qubit stabilisers. Each module applies a local \cnot{} gate between the \network{} and \data{} qubits, mediated temporarily by the \auxiliary{} qubit, mapping the joint \data{}-qubit parity onto the \network{} qubits. Their mid-circuit measurement outcomes are exchanged in real time via a classical (TTL) link.
    \textbf{c}, Characterisation of the two-qubit remote stabiliser measurement. The \data{} qubits are prepared in the four $X$-basis product states before an $\XX$ stabiliser measurement. The probability of measuring odd parity on the \network{}-qubit Bell pair is plotted as a function of the phase, $\phi$, of the final analysis pulse applied to both \network{} qubits before measurement. The $+1$ ($\ket{++},\ket{--}$) and $-1$ ($\ket{+-},\ket{-+}$) stabiliser eigenspaces oscillate $\pi$ radians out of phase, demonstrating syndrome discrimination.
    Error bars are the half-widths of $\SI{68}{\percent}$ binomial confidence intervals, and solid lines are sinusoidal fits to the data.
    }
    \label{fig:XXstabiliser}
\end{figure*}

Photonic interconnects offer a route to this long-range connectivity while simultaneously providing the modularity needed for hardware scaling.
Remote entanglement mediates stabiliser measurements between spatially separated qubits, supporting the implementation of high-rate \ac{qLDPC} codes~\cite{cohen2022lowoverhead, strikis2023quantum,chandra2026distributed}.
\fig{fig:XXstabiliser}(a) illustrates this for a bivariate bicycle code~\cite{bravyi2024high}; partitioning the code across multiple processors reduces the number of qubits and connectivity that must be supported by any individual processor, while remote stabiliser measurements preserve the code connectivity, allowing the high encoding rate of the original code to be retained.
More generally, remote stabiliser measurements provide a primitive for fault-tolerant operations between separately encoded logical qubits, including lattice surgery and related code-deformation protocols~\cite{jacinto2026network,haug2025lattice}.
Substantial progress has been made independently in quantum networking and local \ac{QEC}.
Remote entanglement, teleported quantum gates and distributed algorithms have been demonstrated between networked quantum processors~\cite{qiu2025deterministic, main2025distributed, iuliano2026unconditionally}.
In parallel, experiments within individual devices have demonstrated increasingly powerful error-correction capabilities~\cite{rosenblum2018faulttolerant, ryan-anderson2021realization, bluvstein2024logical, perlin2026faulttolerant}, including logical encodings whose performance improves with increasing code size and demonstrations beyond break-even~\cite{sivak2023realtime,gupta2024encoding,acharya2023suppressing,brock2025quantum}.
However, these capabilities have not previously been combined to extract and use error syndromes for remote \data{} qubits.
Consequently, neither quantum error detection nor correction has been demonstrated across separate quantum processors.
Demonstrating remote syndrome extraction therefore represents a key experimental milestone towards modular fault-tolerant quantum computing.
Such a demonstration tests the complete interface between quantum networking and \ac{QEC}, requiring remote entanglement generation and transfer between \network{} and \data{} qubits, measurement without disturbing the encoded state, and classical communication and feedforward to convert the measurement outcome into an error-detection or correction operation.
Trapped ions provide a natural platform for this modular architecture, combining high-fidelity local quantum logic~\cite{hughes2025trappedion, smith2025singlequbit}, long-lived qubit memories~\cite{langer2005longlived,wang2017singlequbit,sepiol2019probing}, and optical interfaces for remote entanglement generation~\cite{moehring2007entanglement,krutyanskiy2023entanglement}.
Recent demonstrations of gate teleportation between remote trapped-ion processors have established the essential ingredients for deterministic and universal \ac{DQC}~\cite{main2025distributed}, making this platform particularly well suited for exploring networked quantum processors.
Here, we report the first experimental demonstration of quantum error detection and correction across separate quantum processors, enabled by remote syndrome extraction. 
This is done in two complementary experiments using a distributed trapped-ion quantum computer comprising two nodes separated by $\sim\SI{2}{\meter}$.
First, we stabilise the logical subspace of a distributed $[[2,1,1]]$ repetition code by repeatedly measuring the remote stabiliser with ancillary entangled pairs. 
Each syndrome outcome determines whether the run is aborted and restarted, or whether the encoded state is kept for a subsequent round, allowing the protocol to continue until an error is detected. 
We show that this repeated error-detection procedure improves the measured stabiliser with no measurable increase in logical error rate. 
Second, we stabilise a specific entangled Bell state shared between the two processors: using additional entangled pairs, we remotely extract syndromes for arbitrary single-qubit Pauli errors and actively correct them.
These results demonstrate the experimentally distinct primitive required for fault-tolerant \ac{DQC}, namely syndrome extraction for data qubits separated across processors, mediated by entanglement between network qubits.
This provides a proof of principle for remote stabiliser measurements and establishes a route towards full fault-tolerant distributed \ac{QEC}, in which arbitrary logical quantum information can be protected across a network.
\section{Distributed error detection}
We operate a two-node quantum network in which each node is a mixed-species ion-trap quantum computer, as detailed in~\refcite{main2025distributed}. 
The two modules each trap one \Sr{88} ion and one \Ca{43} ion [see \fig{fig:XXstabiliser}(b)]. 
The \Sr{88} ion provides a network qubit, $\Qnet$, while the hyperfine manifold of the \Ca{43} ion provides both a long-lived \data{} qubit, $\Qdat$, and an auxiliary qubit, $\Qaux$, used to mediate local mixed-species gate operations~\cite{drmota2023robust}.
The remote stabiliser measurement, shown in \fig{fig:XXstabiliser}(b), is performed by first distributing a Bell state between the \network{} qubits using the protocol described in refs.~\cite{feng2003entangling,duan2003efficient,simon2003robust}, which has previously been implemented in our setup~\cite{stephenson2020highrate, main2025distributed}.
In brief, each \Sr{88} ion is simultaneously excited to emit a single photon which is polarisation-entangled with its internal spin state.
Interference of two of these single photons on a Bell-state analyser projects the ions onto a Bell state, which we map to $\ket{\Phi^{+}} = \frac{1}{\sqrt{2}}(\ket{00}+\ket{11})$ using local single-qubit gates.
This shared Bell pair between the \network{} qubits acts as a distributed ancilla, allowing the parity of spatially separated \data{} qubits to be measured using only local operations and classical communication.
After heralding remote entanglement between the \network{} qubits, each node applies a local \cnot{} gate between the \network{} and \data{} qubits.
The local two-qubit operations required, here \cnot{} and \iswap{}, are constructed from a universal gate set composed from a mixed-species \ac{CZ} gate and single-qubit rotations. 
We implement the \ac{CZ} gate using a mixed-species geometric phase gate based on a spin-dependent optical dipole force~\cite{cirac1995quantum, hughes2020benchmarking}. 
Measuring the \network{} qubits in the appropriate basis then reveals the $\XX$ (or $\ZZ$) eigenvalue of the remote \data{} qubits.
\begin{figure*}[t]
    \centering
    \includegraphics[width=178mm]{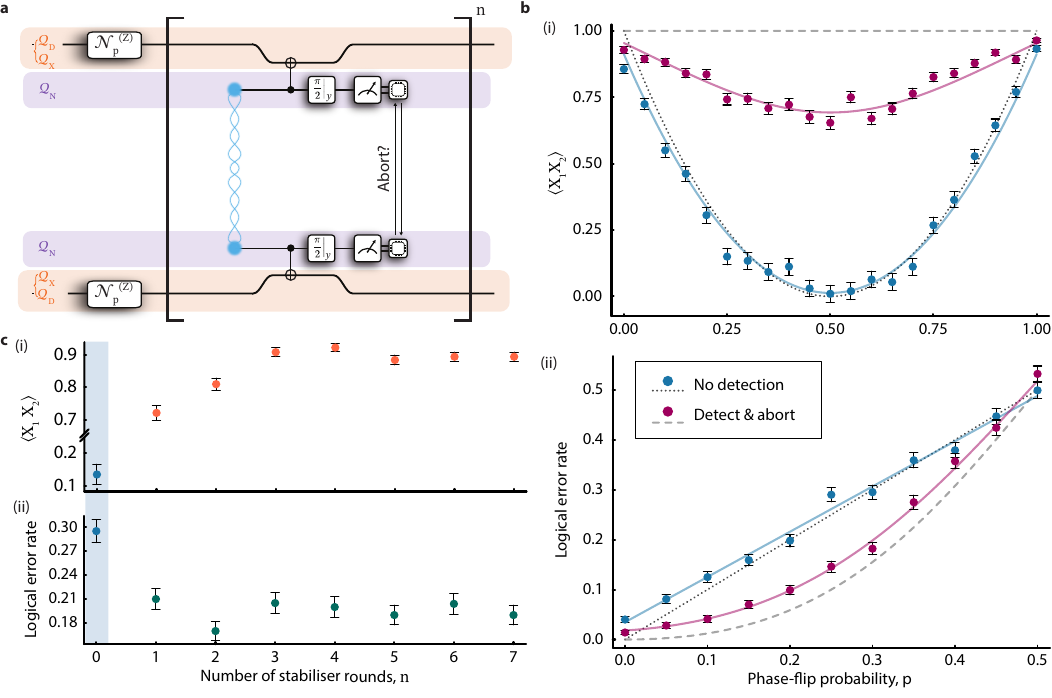}
    \caption{%
    \justifying
    \textbf{Distributed error detection.}
    \textbf{a}, The remote \data{} qubits encode a logical qubit in the repetition-code basis $\ket{0_{\mathrm L}} = \ket{++}$ and $\ket{1_{\mathrm L}} = \ket{--}$, stabilised by $\XX$. $\ket{0_{\mathrm L}}$ is prepared, and each \data{} qubit undergoes an independent phase-flip error with probability $p$. Following a remote $\XX$ stabiliser measurement, an odd-parity outcome causes the logical qubit to be re-prepared and the sequence repeated. This continues until $n$ successive even-parity outcomes are obtained before logical measurement.
    \textbf{b}, Measured \data{}-qubit observables \textbf{i},$\langle \XX \rangle$ and \textbf{ii}, logical error probability as a function of physical phase-flip error probability for $n=0$ (blue) and $n=1$ (pink). The dotted and dashed lines show the corresponding ideal curves without error detection and with an ideal stabiliser measurement, respectively (see Methods).
    \textbf{c}, Measured \data{}-qubit observables \textbf{i}, $\langle \XX \rangle$ and \textbf{ii}, logical error probability after repeated remote stabiliser measurements with $p=0.3$. Increasing $\langle \XX \rangle$ while maintaining an approximately constant logical error rate demonstrates that repeated measurements preserve the code space without introducing additional logical errors.
    Error bars are the half-widths of $\SI{68}{\percent}$ binomial confidence intervals, and solid lines are fits to a model of imperfect stabiliser measurements (see Methods).
    }
    \label{fig:detection}
\end{figure*}
We verify the remote stabiliser measurement in \fig{fig:XXstabiliser}(c) by preparing the \data{} qubits in each of the four two-qubit product states, $\ket{++}$, $\ket{+-}$, $\ket{-+}$, and $\ket{--}$, where $\ket{+}$ and $\ket{-}$ denote the $+1$ and $-1$ eigenstates of $X$, respectively, and applying the remote $\XX$ stabiliser measurement shown in \fig{fig:XXstabiliser}(b).
Each data point presented in this work is obtained from $1,000$ experimental shots.
The states $\ket{++}$ and $\ket{--}$ are $+1$ eigenstates of $\XX$, while $\ket{+-}$ and $\ket{-+}$ are $-1$ eigenstates.
These two subspaces oscillate $\pi$ radians out of phase with each other as the phase, $\phi$, of the final $\pi/2$ analysis pulse applied to the \network{} qubits is scanned.
We attribute the small phase offsets between $\ket{++}$ and $\ket{--}$, and $\ket{+-}$ and $\ket{-+}$ to miscalibrated \ac{CZ} gates (see Methods).
The observed $\pi$ radians separation confirms that the measurement distinguishes the two stabiliser subspaces.
To demonstrate this remote stabiliser measurement, we consider the two-qubit repetition code encoded in the $X$ basis, 
\begin{equation}
    \ket{0_{\mathrm L}} = \ket{++}, \qquad
    \ket{1_{\mathrm L}} = \ket{--}.
\end{equation}
The corresponding logical operators are chosen as $X_{\mathrm L} = \ZZ$ and $Z_{\mathrm L} =\IX$. 
We choose this encoding because the dominant noise mechanism for trapped-ion qubits is dephasing due to magnetic-field fluctuations, whereas bit-flip errors are comparatively rare owing to long population lifetimes, high-fidelity state preparation and readout~\cite{sotirova2024highfidelity}, and high-fidelity single-qubit operations~\cite{smith2025singlequbit}.
A phase-flip error maps the encoded state out of the code space and inverts the $\XX$ parity, allowing the error to be detected.
We tune the phase-flip error rate by probabilistically applying physical $Z$-rotations to the \data{} qubits (see Methods), realising, over many repetitions, an independent phase-flip channel with probability $p$ on each \data{} qubit.
This allows us to demonstrate how the remote stabiliser measurement suppresses the dominant error channel in our system.
We prepare the logical state $\ket{0_{\mathrm L}}$ and subject it to the phase-flip channel before applying the remote $\XX$ stabiliser measurement.
Whenever the syndrome inferred from the \network{}-qubit measurement records an error ($\XX= -1$), the attempt is aborted in real time, the logical qubit is reinitialised, and the sequence is repeated until the \network{} qubit measurement indicates no error ($\XX = +1$). 
The final state of the logical qubit is then characterised by measuring the \data{}-qubit observables $\langle \XX \rangle$ (which measures occupation of the code space) and $\langle Z_{\mathrm L} \rangle = \langle \IX \rangle$, to infer the logical error probability, $P_{\mathrm L} = (1-\langle Z_{\mathrm L} \rangle)/2$.
As shown in \fig{fig:detection}(b), conditioning on the remote stabiliser outcome both increases the probability that the logical qubit remains in the code space and suppresses logical errors across the full range of applied error rates. 
At zero applied error, the probability of obtaining the target $\ket{++}$ outcome after the stabiliser measurement is $P_{++}=\Ppp$, corresponding to an error of $1-P_{++}=\Ppperror$. 
The independently characterised contributions to this error are summarised in Table~\ref{tab:errorbudget} and described in Methods.
\begin{table}[t]
    \centering
    \include{figures/error_budget.tex}
    \caption{%
    \justifying
    \textbf{Error budget for one $\XX$ stabiliser measurement.}
    Independently characterised errors and their modelled contributions to the measured $P_{++}$ error.
    }
    \label{tab:errorbudget}
\end{table}
Since the coherence time of the \data{} qubits is orders of magnitude greater than the time it takes to generate another \network{}-qubit Bell state (approximately \SI{10}{\second} and \SI{100}{\milli\second}, respectively~\cite{main2025multipartite}), we are able to perform multiple rounds of stabiliser measurements.
This is shown in \fig{fig:detection}(c), where we prepare the logical state $\ket{0_{{\mathrm L}}}$ and subject it to the phase-flip channel with $p=0.3$.
The remote $\XX$ stabiliser is then measured repeatedly, aborting the protocol in real time and restarting when a phase-flip error is detected.
Because the stabiliser measurement is imperfect, some single-qubit errors remain undetected after a round. 
Successive conditioning progressively filters out these errors, increasing $\langle\XX\rangle$ and hence the confidence that the state lies within the code space.
We also show that the logical error rate remains approximately constant with repeated measurements, confirming that the encoded information is preserved in the process.
This demonstrates that the remote stabiliser measurement can be performed repeatedly and used to continuously suppress phase errors in a delocalised logical qubit.
\section{Error correction of a Bell state}
\begin{figure*}[t]
    \centering
    \includegraphics[width=178mm]{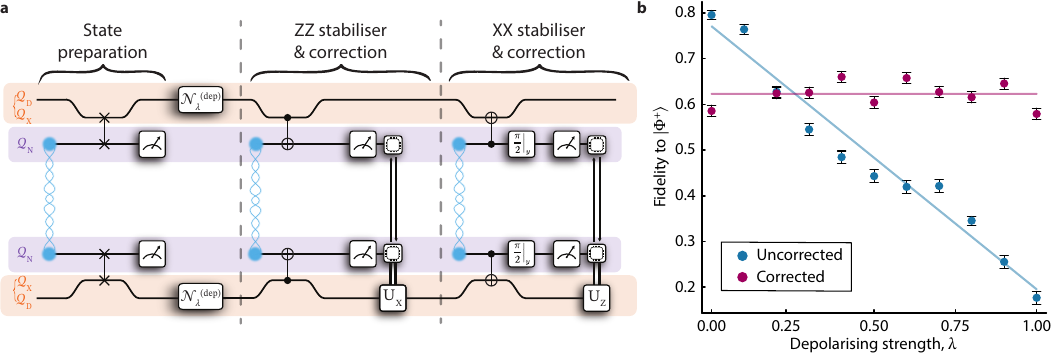}
    \caption{%
    \justifying
    \textbf{Error correction of a Bell state.}
    \textbf{a}, The \data{} qubits are prepared in the Bell state $\ket{\Phi^{+}}$ by mapping from a heralded \network{} pair using local error-detected \iswap{} gates. This state undergoes injected depolarising noise with strength $\lambda$ before syndrome extraction is performed with both $\ZZ$ and $\XX$ stabilisers. Detected bit- and phase-flip errors are corrected by applying $U_{\mathrm X}$ and $U_{\mathrm Z}$, respectively.
    \textbf{b}, By measuring both $\ZZ$ and $\XX$ stabilisers and applying the conditioned corrections, the fidelity to $\ket{\Phi^{+}}$ remains approximately constant across the range of injected error strengths. Error bars are the half-widths of $\SI{68}{\percent}$ binomial confidence intervals, and solid lines are linear and constant fits to the uncorrected and corrected data, respectively.
    }
    \label{fig:bellcorrection}
\end{figure*}
Practical \ac{QEC} ultimately requires measured syndromes to be used for feedforward correction, while preserving encoded quantum information that will generally exist in a coherent superposition.
As \ac{QEC} of arbitrary logical superpositions requires at least 5 physical qubits, here we consider protecting specifically the Bell state $\ket{\Phi^{+}} = \frac{1}{\sqrt{2}}(\ket{00} + \ket{11})$, which is stabilised by both $\XX$ and $\ZZ$.
This allows us to demonstrate active correction of both phase-flip and bit-flip errors using the measured remote syndromes.
The measured syndrome uniquely determines the required recovery operation; a single error on either \data{} qubit transforms $\ket{\Phi^{+}}$ into the same orthogonal Bell state, allowing the correction to be applied in either node to restore $\ket{\Phi^{+}}$.
While a known Bell state could be re-prepared instead of corrected, preserving it through active feedforward provides a simple testbed for demonstrating the complete remote error-correction cycle, including syndrome extraction, classical feedforward, and conditional correction.
Extending this approach to protect non-trivial logical qubits requires encoding more physical qubits than the number of independent stabiliser generators.
To prepare the remote \data{} qubits in $\ket{\Phi^{+}}$, we first generate entanglement between the \network{} qubits, which we then map to the \data{} qubits using local error-detected \iswap{} gates in each node~\cite{drmota2023robust} [see \fig{fig:bellcorrection}(a)].
This prepared Bell state has an average fidelity of \cacarawfidelity{}.
Depolarising noise is then injected into this state (see Methods), described by the channel
\begin{equation}
    \mathcal{N}_{\mathrm{\lambda}}^{\mathrm{(dep)}}(\rho)=\left(1-\frac{3\lambda}{4}\right)\rho +
\frac{\lambda}{4}\left(X\rho X+Y\rho Y+Z\rho Z\right),
\label{eq:depolarising channel}
\end{equation}
where $X$, $Y$, and $Z$ are the single-qubit Pauli gates and $\lambda \in [0,1]$ parametrises the noise strength such that, when $\lambda=1$, the state becomes maximally mixed. 
The depolarising channel acts on both \data{} qubits independently.
We then perform sequential remote measurements of both the $\ZZ$ and $\XX$ stabilisers, as shown in \fig{fig:bellcorrection}(a), to detect bit-flip and phase-flip errors, respectively.
Upon detection of a bit-flip error, we apply an $X$ rotation to Bob's \data{} qubit as a real-time correction.
We apply an explicit physical rotation here, although the same operation can be done virtually by updating the measurement bases, avoiding additional experimental delays.
Upon detection of a phase-flip error, the correction is implemented as a virtual $Z$ rotation by advancing the phase reference of Bob's \data{} qubit by $\pi$ radians.
Each experimental shot requires three rounds of remote entanglement generation: one to prepare the \data{} qubits in $\ket{\Phi^{+}}$, and one for each of the $\ZZ$ and $\XX$ remote stabilisers.
In \fig{fig:bellcorrection}(b) we compare the estimated fidelity of the two-\data{}-qubit state to $\ket{\Phi^{+}}$ with and without remote stabiliser measurements and correction for the same value of the depolarising parameter $\lambda$. 
The fidelity to the ideal state $\ket{\Phi^{+}}$ is estimated from measurements of the population and coherence using
\begin{equation}
    F = \frac{1+\langle \XX \rangle - \langle \YY \rangle + \langle \ZZ\rangle}{4}.
\label{eq:fidelimetry}
\end{equation}
When applying corrections conditioned on the remote $\XX$ and $\ZZ$ stabiliser measurements, the fidelity to $\ket{\Phi^{+}}$ remains approximately constant across the applied range of depolarising noise.
In contrast, the fidelity decreases monotonically with depolarising strength when we measure directly after preparing the noisy state.
The lower fidelity at $\lambda = 0$ reflects the finite infidelity of the stabiliser measurements, which introduces a fixed experimental overhead that is independent of depolarising noise (see Methods).
By combining remote $\XX$ and $\ZZ$ stabiliser measurements, we detect and correct arbitrary single-qubit Pauli errors on a distributed entangled state.
These results demonstrate the first active correction of errors using remotely measured stabilisers, establishing a key building block for implementing \ac{QEC} across networked quantum computing modules.

\section{Discussion}
Our results establish remote syndrome extraction as an experimental primitive for distributed \ac{QEC}. 
By using entanglement shared between two physically separated processors to measure and act on remote stabilisers, we show that such error-detection and correction protocols can be extended beyond a single quantum device. 
The experiment combines remote entanglement generation, local quantum logic, local and non-local mid-circuit measurement, real-time classical communication between the processors, and feedforward in a single error-correction workflow.
This demonstrates that all the essential ingredients for \ac{QEC} can operate coherently and repeatedly across a quantum network.
The fidelities achieved are not sufficient for typical fault-tolerant implementations, but they are limited mainly by technical imperfections in the local entangling gates used to interface \network{} and \data{} qubits (see Table~\ref{tab:errorbudget}). 
This is a well-understood challenge in trapped-ion systems, rather than a fundamental limitation of the protocol. 
The same class of gates has already been performed with 99.8(1)\% fidelity using laser-driven interactions~\cite{hughes2020benchmarking}, and trapped-ion entangling gates more broadly have reached fidelities of $>$99.99\% in optimised systems~\cite{hughes2025trappedion}.
Another limitation to the quality of remote stabiliser measurements is the fidelity of the entangled network link.
While our remote entanglement fidelity, together with that demonstrated in~\refcite{oreilly2024fast}, represents the current state of the art, high-fidelity remote Bell pairs are not a strict requirement for scalable distributed \ac{QEC}.
Certain fault-tolerant distributed architectures can tolerate significantly noisier inter-module Bell pairs, provided that local operations retain sufficiently high fidelity, and the network errors are incorporated into the error-correction protocol~\cite{ramette2024faulttolerant, nickerson2013topological}.
Entanglement purification, where multiple low fidelity remote Bell pairs are converted into a smaller number of higher fidelity pairs, can also be used if higher fidelity network entanglement is required~\cite{dur2003entanglement,nigmatullin2016minimally}.
Bell-state stabilisation can then itself be nested within such purification protocols for further error suppression towards the level of local operations.
Since remote stabiliser measurements require the prior generation of a distributed Bell pair, network latency can become a dominant contribution to the error correction cycle time.
During this extended idle time, the \data{} qubits will continue to accumulate errors, reducing the effectiveness of the syndrome extraction~\cite{kaur2026impact, singh2025modular}.
Several approaches to minimise this bottleneck have been proposed, including continuous Bell-state generation to reduce idling time~\cite{fu2025efficient}, optimised stabiliser scheduling that overlaps communication with local operations~\cite{gupta2026boundaryaware}, and distributed \ac{QEC} codes that are more robust to network latencies by minimising the number of remote stabiliser measurements~\cite{sutcliffe2025distributed, galimova2026distributed} that are required.
In our setup, remote Bell states are generated at an average rate of approximately \SI{10}{\second^{-1}} when interleaved with local entangling operations, including additional cooling of the \Sr{88}-\Ca{43} crystal (see Methods).
Future improvements could increase this rate through cavity-enhanced entanglement efficiency~\cite{krutyanskiy2023entanglement} and ion-trap architectures that enable continuous sympathetic cooling without interrupting the protocol~\cite{oreilly2024fast}.
These results are complementary to recent progress in local trapped-ion \ac{QEC}.
A scalable trapped-ion quantum computer could combine high-fidelity local \ac{QEC}, for example within QCCD-style modules~\cite{kielpinski2002architecture,paetznick2026improved}, with remote stabiliser measurements between modules. 
The same principle can be applied to any quantum computing platform capable of generating entanglement between distant nodes and coupling it to local data qubits, including neutral atoms~\cite{reiserer2015cavitybased}, colour centres~\cite{knaut2024entanglement, humphreys2018deterministic, inc2024distributed}, superconducting circuits~\cite{storz2023loopholefree, axline2018ondemand}, and other emergent modular architectures. 
Remote syndrome extraction may also be useful beyond quantum computing, for example, in protecting distributed entangled states for quantum communication~\cite{muralidharan2014ultrafast}, and networked sensing and metrology~\cite{dur2014improved,arrad2014increasing}. 
By combining remote entanglement, local quantum logic, mid-circuit measurement and feedforward, our work marks a step from scalable distributed quantum operations towards truly fault-tolerant scalable quantum networks.
\input{methods}

\FloatBarrier

\section{Data Availability}
The datasets generated during the current study are available from \ellis{} and \gabriel{} upon reasonable request.

\section{Code Availability}
The analysis code that supports the plots in this paper and the other findings of this study is available from \ellis{} upon reasonable request.

\section*{Rights Retention}
This research was funded in whole, or in part, by a Plan S funder. For the purpose of Open Access, the author has applied a CC BY public copyright licence to any Author Accepted Manuscript version arising from this submission.
\input{acronyms.tex}
\bibliography{library}
\input{end_notes.tex}
\end{document}

%% file: style.tex
\usepackage{graphicx}
\usepackage{caption}
\usepackage{ragged2e}
\usepackage{bm}
\usepackage{svg}
\usepackage{siunitx}
\usepackage{amsmath}
\usepackage{amssymb}
\usepackage[nolist]{acronym}
\usepackage{tabularray}
\UseTblrLibrary{booktabs}
\usepackage{csquotes}
\usepackage{physics}
\usepackage{pifont, xcolor}
\usepackage{placeins}
\usepackage{afterpage}

\usepackage[hidelinks]{hyperref}

\setcitestyle{super, sort&compress}
\newcommand{\refcite}[1]{Ref.~\citenum{#1}}

%% file: commands.tex
\newcommand{\Title}{Error Correction in a Distributed Quantum Computer}

\newcommand{\fig}[1]{Fig.~\ref{#1}}

\newcommand{\methods}[1]{\hyperref[#1]{(Methods)}}

\newcommand{\iswap}{iSWAP}
\newcommand{\cnot}{CNOT}

\newcommand{\network}{network}
\newcommand{\auxiliary}{auxiliary}
\newcommand{\data}{data}

\newcommand{\ion}[2]{\mbox{$^{#2}$#1$^+$}}
\newcommand{\Ca}[1]{\ion{Ca}{#1}}
\newcommand{\Sr}[1]{\ion{Sr}{#1}}

\newcommand{\Q}[1]{\mathcal{Q}_{#1}}
\newcommand{\net}{\mathrm{N}}
\newcommand{\Qnet}{\Q{\net{}}}
\newcommand{\aux}{\mathrm{X}}
\newcommand{\Qaux}{\Q{\aux{}}}
\newcommand{\dat}{\mathrm{D}}
\newcommand{\Qdat}{\Q{\dat{}}}

\newcommand{\XX}{X_1 X_2}  
\newcommand{\YY}{Y_1 Y_2}  
\newcommand{\ZZ}{Z_1 Z_2}  
\newcommand{\IX}{I_1 X_2}  

\newcommand{\cacarawfidelity}{\SI{82.5(15)}{\percent}} 

\newcommand{\Ppp}{\SI{96.4(6)}{\percent}}
\newcommand{\Ppperror}{\SI{3.6(6)}{\percent}}

\newcommand{\rawinfidelity}{\SI{2.38(17)}{\percent}}
\newcommand{\rawinfidelitycontribution}{\SI{0.16(1)}{\percent}}

\newcommand{\alicememoryerrors}{\SI{1.4(3)}{\percent}}
\newcommand{\alicememoryerrorscontribution}{\SI{0.15(3)}{\percent}}
\newcommand{\bobmemoryerrors}{\SI{5.9(3)}{\percent}}
\newcommand{\bobmemoryerrorscontribution}{\SI{0.41(2)}{\percent}}

\newcommand{\alicewzz}{\SI{3.43(9)}{\percent}}
\newcommand{\alicewzzcontribution}{\SI{1.35(4)}{\percent}}

\newcommand{\bobwzz}{\SI{3.04(9)}{\percent}} 
\newcommand{\bobwzzcontribution}{\SI{1.15(3)}{\percent}} 

\newcommand{\alicetransfer}{\SI{0.94(13)}{\percent}}
\newcommand{\alicetransfercontribution}{\SI{0.50(7)}{\percent}} 
\newcommand{\bobtransfer}{\SI{0.21(2)}{\percent}} 
\newcommand{\bobtransfercontribution}{\SI{0.11(1)}{\percent}} 

\newcommand{\alicemidcircuiterror}{\SI{0.113(3)}{\percent}}
\newcommand{\alicemidcircuiterrorcontribution}{\SI{0.0112(3)}{\percent}}
\newcommand{\bobmidcircuiterror}{\SI{0.084(1)}{\percent}} 
\newcommand{\bobmidcircuiterrorcontribution}{\SI{0.0083(1)}{\percent}}

\newcommand{\aliceclockrotations}{\SI{0.107(2)}{\percent}}
\newcommand{\aliceclockrotationscontribution}{\SI{0.063(1)}{\percent}}
\newcommand{\bobclockrotations}{\SI{0.136(1)}{\percent}}
\newcommand{\bobclockrotationscontribution}{\SI{0.078(1)}{\percent}}%

\newcommand{\totalerrorcontribution}{\SI{3.5(2)}{\percent}} %

\newcommand{\ellis}{EMA}
\newcommand{\ayush}{AA}
\newcommand{\tenzan}{TA}
\newcommand{\dougal}{DM}
\newcommand{\adam}{ARM}
\newcommand{\erin}{EM}
\newcommand{\marion}{MM}
\newcommand{\shuying}{SC}
\newcommand{\jake}{JAB}
\newcommand{\davidN}{DPN}
\newcommand{\peter}{PD}
\newcommand{\raghu}{RS}
\newcommand{\simon}{SCB}
\newcommand{\davidL}{DML}
\newcommand{\gabriel}{GA}

%% file: title-config.tex
\author{E.~M.~Ainley}
\email{ellis.ainley@physics.ox.ac.uk}
\author{A.~Agrawal}
\affiliation{Department of Physics, University of Oxford, Clarendon Laboratory, Parks Road, Oxford OX1 3PU, United Kingdom}
\author{T.~Araki}
\affiliation{Department of Physics, University of Oxford, Clarendon Laboratory, Parks Road, Oxford OX1 3PU, United Kingdom}
\affiliation{Mathematical Institute, University of Oxford, Woodstock Road, Oxford OX2 6GG, United Kingdom}
\author{A.~R.~Mart\'{\i}nez}
\author{D.~Main}
\author{E.~Malinowski}
\author{J.~A.~Blackmore}
\author{S.~Chen}
\author{P.~Drmota}
\author{M.~Mallweger}
\author{D.~P.~Nadlinger}
\author{R.~Srinivas}
\affiliation{Department of Physics, University of Oxford, Clarendon Laboratory, Parks Road, Oxford OX1 3PU, United Kingdom}
\author{S.~C.~Benjamin}
\affiliation{Department of Materials, University of Oxford, Parks Road, Oxford OX1 3PH, United Kingdom}
\author{G.~Araneda}
\email{gabriel.aranedamachuca@physics.ox.ac.uk}
\author{D.~M.~Lucas}
\affiliation{Department of Physics, University of Oxford, Clarendon Laboratory, Parks Road, Oxford OX1 3PU, United Kingdom}

%% file: figures/error_budget.tex
\begin{tblr}{
    width = \columnwidth,
    colspec = {X[l,1.5] Q[l,wd=0.9cm] X[c,0.9] X[c,1]},
    column{2} = {leftsep=0pt, rightsep=2pt},
}
    \hline[2pt]
    \SetCell[c=2]{l} Source
        & & Raw error & Contribution to $1 - P_{++}$ \\
    \hline[1pt]

    \SetCell[c=2]{l} Raw entanglement
        & & \rawinfidelity{}
        & \rawinfidelitycontribution{} \\

    \SetCell[r=2]{l,m} Mixed-species gate
        & Alice
        & \alicewzz{}
        & \alicewzzcontribution{} \\
        & Bob
        & \bobwzz{}
        & \bobwzzcontribution{} \\

    \SetCell[r=2]{l,m} $\Q{D}$ decoherence
        & Alice
        & \alicememoryerrors
        & \alicememoryerrorscontribution \\
        & Bob
        & \bobmemoryerrors
        & \bobmemoryerrorscontribution \\

    \SetCell[r=2]{l,m} $\Qaux \leftrightarrow \Qdat$ transfer
        & Alice
        & \alicetransfer
        & \alicetransfercontribution \\
        & Bob
        & \bobtransfer
        & \bobtransfercontribution \\

    \SetCell[r=2]{l,m} $\Qnet$ measurement
        & Alice
        & \alicemidcircuiterror
        & \alicemidcircuiterrorcontribution \\
        & Bob
        & \bobmidcircuiterror
        & \bobmidcircuiterrorcontribution \\

    \SetCell[r=2]{l,m} $\Qdat{}$ rotations
        & Alice
        & \aliceclockrotations
        & \aliceclockrotationscontribution \\
        & Bob
        & \bobclockrotations
        & \bobclockrotationscontribution \\

    \hline[1pt]
    \SetCell[c=2]{l} Predicted total error
        & & & \totalerrorcontribution{} \\
    
    \SetCell[c=2]{l} Measured total error
        & & & \Ppperror{} \\
    \hline[2pt]
\end{tblr}

%% file: methods.tex
\section{Methods}
\subsection{Effect of coherent CZ gate errors on the stabiliser measurement}
The remote $\XX$ stabiliser measurement is implemented using local \cnot{} gates with the \network{} qubits as the controls and the \data{} qubits as the targets. 
Ideally, the two-qubit product states map the network Bell state to
\[
\ket{\Phi(\Theta)}=\frac{1}{\sqrt{2}}\left(\ket{00}+e^{i\Theta}\ket{11}\right),
\]
with
\begin{align*}
\ket{++},\ket{--} &: \Theta=0 \quad (\ket{\Phi^+}),\\
\ket{+-},\ket{-+} &: \Theta=\pi \quad (\ket{\Phi^-}),
\end{align*}
such that the two $\XX$ eigenspaces are distinguished by a relative Bell-state phase of $\pi$.

We model residual coherent calibration errors in the local mapping operations by introducing effective conditional phase offsets $\delta_A$ and $\delta_B$ for Alice and Bob's \cnot{} gates, respectively. 
Such offsets may arise from coherent miscalibration of the local \ac{CZ} gate, for example due to laser pointing drifts or imperfect initial calibration resulting in an implemented controlled phase of $\pi+\delta$ rather than the ideal value $\pi$.
The resulting Bell-state phases are
\begin{align*}
\Theta_{++} &= 0,\\
\Theta_{+-} &= \pi+\delta_B,\\
\Theta_{-+} &= \pi+\delta_A,\\
\Theta_{--} &=\delta_A+\delta_B.
\end{align*}

The centres of the two stabiliser eigenspaces therefore remain separated by $\pi$, while the individual states within each eigenspace acquire relative phase shifts
\begin{align*}
\Delta_{+}&=\Theta_{--}-\Theta_{++}=\delta_{\mathrm A}+\delta_{\mathrm B},\\
\Delta_{-}&=\Theta_{-+}-\Theta_{+-}=\delta_{\mathrm A}-\delta_{\mathrm B}.
\end{align*}

These coherent offsets shift the phase of the measured \network{}-qubit parity fringes while preserving the parity-dependent inversion of the stabiliser signal. 
If uncorrected, they reduce the discrimination between the two stabiliser eigenspaces and therefore increase the probability of incorrect syndrome assignment.
In principle, these offsets can be calibrated by measuring the phase of the parity fringes and compensated by applying appropriate phase corrections to the local mapping operations. 
No such compensation was applied in the experiments reported here.

\subsection{Phase-flip error channel}
Throughout an experimental shot, a tracked phase, $\varphi$, is maintained in software for each qubit, representing the qubit's phase in the rotating frame.
$\varphi$ is updated to account for any deterministic phase offsets introduced by single-qubit rotations, entangling gates, and the known phase associated with the heralded Bell-state outcome.
This ensures that all subsequent operations, including the final analysis pulse, are applied about the intended axis.
For each coherent operation, $\varphi$ is added to the phase of the qubit's driving laser by programming the phase of the \ac{DDS} chip driving the corresponding \ac{AOM}.
The accumulated tracked phase can then be transferred from the software to the physical qubit by applying two resonant $\pi$-pulses with phases $\pm\varphi/2$,
\[
R_{\varphi/2}(\pi)R_{-\varphi/2}(\pi) = e^{-i\varphi Z/2},
\]
such that the final analysis pulse, and hence the measurement basis, is defined with respect to the tracked reference frame.
Artificial phase-flip errors are implemented by applying a physical $Z$-rotation to the qubits.
For each experimental repetition, each module independently generates a pseudo-random number to determine whether a phase flip is applied to its \data{} qubit with probability $p$.
If yes, a $Z(\pi)$ rotation is applied:
\[
R_{\pi/2}(\pi)R_{-\pi/2}(\pi) = Z,
\]
but unlike deterministic phase rotations, the tracked phase is intentionally not updated.
Subsequent coherent operations are performed relative to the original reference frame, realising the effect of a physical phase-flip error.
Repeating this procedure over many shots implements the phase-flip channel
\[
    \mathcal{N}_{\mathrm p}^{\mathrm Z}(\rho)=\left(1-p\right)\rho + pZ\rho Z.
\]

\subsection{Depolarising error channel}
As with the phase-flip channel, Alice and Bob independently determine whether to apply an error to their respective \data{} qubits with probability $\lambda$.
If selected, one single-qubit Pauli operation $\{I,X,Y,Z\}$ is chosen uniformly at random.
$X$ is implemented with a $\pi$-pulse on the qubit, $Z$ as above, and $Y$ by applying an $X$ operation and then a $Z$ operation.
These operations are applied physically and are not added to the tracked phase reference.
Averaging over many shots realises the depolarising channel given in Eq.~\eqref{eq:depolarising channel}.

\subsection{Model of stabiliser measurements}
To model the effect of the conditioned stabiliser measurements, we first assume each \data{} qubit experiences an independent phase-flip error with probability $p$.
The probabilities of zero, one, and two phase-flip errors are then given by 
\begin{align*}
P_{0} &= (1-p)^2, \\
P_{1} &= 2p(1-p), \\
P_{2} &= p^2
\end{align*}
To account for imperfect stabiliser performance, we introduce acceptance probabilities $r_0$, $r_1$, and $r_2$ for events containing zero, one, and two physical errors, respectively.
The total acceptance probability is therefore
\[
S = r_0 P_0 + r_1 P_1 + r_2 P_2.
\]
Since the conditioned observables are invariant under an overall scaling of $r_0$, $r_1$, $r_2$, the model constrains only the relative values $r_1/r_0$ and $r_2/r_0$.

To model the $\langle \XX \rangle$ measurement, single phase-flip errors invert the parity, whereas zero and two phase-flip errors preserve it. 
Without stabiliser conditioning, 
\[
\langle \XX\rangle=P_0-P_1+P_2=(1-2p)^2.
\]
Conditioning on the remote stabiliser measurement gives
\[
\langle \XX\rangle_{\mathrm{fit}} = C+B\frac{r_0 P_0-r_1P_1+r_2P_2}{S},
\]
where $B$ and $C$ account for non-unity contrast and residual \ac{SPAM} errors, respectively.
In the ideal limit this reduces to $\langle \XX\rangle=1$.

For the logical observable $\langle Z_{\mathrm L}\rangle=\langle \IX \rangle$, zero- and two-error events contribute $+1$ and $-1$, respectively. 
We assume accepted single-error events contribute equally to the $\pm1$ outcomes of $\langle Z_{\mathrm L}\rangle$, and therefore make no net contribution to the expectation value. 
The fitted model is therefore
\[
\langle Z_{\mathrm L}\rangle_{\mathrm{fit}} = C+B\frac{r_0 P_0-r_2P_2}{S},
\]
with the corresponding logical error probability
\[
P_{\mathrm L}=\frac{1-\langle Z_{\mathrm L}\rangle_{\mathrm{fit}}}{2}.
\]

\subsection{Cooling and rates}
The mixed species geometric phase gate used in the experiments is implemented by driving the axial \ac{OOP} motional mode with a spin-dependant force.
This gate is intrinsically insensitive to the initial occupation of the \ac{OOP} mode of the two-ion chain, given that the ions remain in the Lamb-Dicke regime. 
Excitation of the other motional modes of the \Sr{88}-\Ca{43} crystal can also reduce the gate fidelity through several mechanisms.
The ratio of the axial \ac{OOP} and axial \ac{IP} mode frequencies is close to 2, placing the second harmonic of the \ac{IP} mode close to the \ac{OOP} mode frequency. 
As a result, the gate drive unavoidably couples to the \ac{IP} motion, leaving residual spin-motion entanglement at the end of the gate, with the resulting error increasing with the occupation of the \ac{IP} mode.
Although they are orthogonal to the gate drive, finite occupation of the radial modes gives rise to anharmonic Coulomb interactions, producing higher-order Kerr cross-couplings and other non-linear mode-mixing effects that perturb the \ac{OOP} mode frequency during the gate.
The probabilistic nature of remote entanglement generation results in long idle periods before each mixed-species gate, during which the motional modes will heat up.
To suppress these error mechanisms, we interleave \SI{200}{\micro\second} of entanglement generation attempts with re-cooling of the \Sr{88}-\Ca{43} crystal. 
The axial \ac{IP} and \ac{OOP} modes are cooled with \SI{700}{\micro\second} of \Sr{88} \ac{EIT} cooling.
The radial modes are re-cooled using \SI{1.5}{\milli\second} of Doppler cooling on the \Sr{88}. 
While the two radial modes with mostly \Sr{88} motion are cooled directly, the remaining two radial modes are dominated by \Ca{43} motion and cannot be cooled directly.
Direct cooling of the \Ca{43} ions would cause photon scattering, destroying the quantum information encoded in their hyperfine qubits.
Instead, we sympathetically cool these modes by modifying the trapping potential to increase the participation of the \Sr{88} ion in the \Ca{43}-dominated radial modes.
This increases the effective Lamb-Dicke parameter for the Doppler cooling transition and improves the cooling efficiency.
This recooling sequence is essential for maintaining the mixed-species gates fidelity. 
However, the additional \SI{2.2}{\milli\second} cooling overhead reduces the remote entanglement generation rate from approximately \SI{100}{\per\second} to approximately \SI{10}{\per\second}.

\subsection{Error budget}
The dominant contributions to the experimental error are summarised in Table~\ref{tab:errorbudget}. 
Each contribution is characterised independently using dedicated calibration experiments and then incorporated into a model of the stabiliser-measurement circuit.
The raw entanglement error is obtained from quantum state tomography of the remotely entangled \network{} qubits.
The mixed-species entangling-gate error is obtained from quantum process tomography.
Decoherence of the \data{} qubits is characterised by measuring the decay of $X$-basis coherence under dynamical decoupling over multiple entanglement-generation rounds, from which the decoherence per round is extracted.
The error associated with \auxiliary{}--\data{} transfer is determined using a modified single-qubit \ac{RBM} sequence that alternates Clifford operations between the two qubit encodings, with the quoted error corresponding to the two transfers used in the stabiliser-measurement circuit.
The mid-circuit measurement error is calculated from the error of the parity-mapping rotation on the \network{} qubit, obtained from single-qubit \ac{RBM}.
Errors in the \data{}-qubit rotations are obtained from single-qubit \ac{RBM}, with the quoted error corresponding to the seven rotations performed during the stabiliser-measurement circuit.
The independently characterised errors are converted to their corresponding noise channels and applied at the appropriate locations in a model of the complete stabiliser-measurement circuit.
The predicted total error is obtained with all characterised error sources included in the model.
The contribution of each error source to the $P_{++}$ error is estimated from the change in $P_{++}$ when that source is removed from the full model, with all other characterised errors retained.
This accounts for the fact that errors occurring at different stages of the circuit do not contribute independently to the final measurement outcome.

This error budget does not include calibration drifts occurring over the duration of the experimental data acquisition, including slow variations in gate times and frequencies, and residual phase offsets in the mixed-species entangling gates described above. 
Such effects are difficult to quantify through independent calibrations, as they accumulate over the course of long experimental runs.

%% file: acronyms.tex
\begin{acronym}
    \acro{QEC}{quantum error correction}
    \acro{qLDPC}{quantum low-density parity-check}
    \acro{CZ}{controlled-$Z$}
    \acro{EIT}{electromagnetically-induced transparency}
    \acro{SDF}{spin-dependant force}
    \acro{DDS}{direct digital synthesis}
    \acro{AOM}{acousto-optic modulator}
    \acro{SPAM}{state preparation and measurement}
    \acro{DQC}{distributed quantum computing}
    \acro{OOP}{out of phase}
    \acro{IP}{in phase}
    \acro{RBM}{randomised benchmarking}
\end{acronym}

%% file: end_notes.tex
\section{Acknowledgements}
We thank Chris Ballance, Bethan Nichol, and Laurent Stephenson for their contributions to the design and construction of the apparatus, Sandia National Laboratories for supplying the ion traps used in this experiment, and the developers of the control system ARTIQ~\cite{ARTIQ}.
\ellis{} acknowledges support from the U.K.\ EPSRC \enquote{Quantum Communications} Hub EP/T001011/1.
\tenzan{} acknowledges support from the Oxford-Uehiro Graduate Scholarship Programme.
\erin{} acknowledges support from the U.K.\ NQCC Doctoral Studentship Scheme.
\adam{} acknowledges support from the Rhodes Trust. 
\jake{} acknowledges support from an \ EPSRC Fellowship UKRI1223 and Reuben College, Oxford.
\davidN{} acknowledges support from Merton College, Oxford.
\raghu{} acknowledges funding from an EPSRC Fellowship EP/W028026/1 and Balliol College, Oxford.
\gabriel{} acknowledges support from Wolfson College Oxford and Cisco.
This work was supported by the U.K.\ EPSRC via the \enquote{Quantum Computing and Simulation} Hub EP/T001062/1, the \enquote{Integrated Quantum Networks} Hub EP/Z533208/1, and the \enquote{Quantum Computing via Integrated and Interconnected Implementations} (QCI$^3$) Hub EP/Y024389/1, and by Innovate UK via UKRI3703.

\section{Author Information}

\subsection{Contributions}
\ellis{}, \ayush{}, \adam{}, \dougal{}{}, \erin{}, \jake{}, \shuying{}, \peter{}, \marion{}, \davidN{}, \raghu{} and \gabriel{} built and operated the experimental apparatus.
\ellis{} led the experimental work, with assistance from \ayush{}.
\ellis{} performed the data analysis and prepared the manuscript.
\ellis{}, \tenzan{}, \simon{} and \gabriel{} conceived the experiments.
\davidL{} and \gabriel{} secured the funding and supervised the work.
All authors contributed to the discussion of the results and manuscript.

\subsection{Competing Interests}
\dougal{} is fully and \raghu{} partially employed by Oxford Ionics Ltd. 
\peter{} is a director of and partially employed by Quantum Fabrix Ltd. 
\simon{} is the founder of Quantum Motion Technologies Ltd.
\davidL{} consults for Quantinuum Ltd.
The remaining authors declare no competing interests.

\subsection{Corresponding Authors}
Correspondence should be addressed to \ellis{} or \gabriel. 